\documentclass[twocolumn,prb,showpacs]{revtex4}
\usepackage{graphicx}
\usepackage{bm}
\newcommand\pyr{pyrochlore}
\newcommand\Ho{Ho$^{3+}$}
\newcommand\Mlll{$M_{111}$}
\newcommand\Mllo{$M_{110}$}
\newcommand\Mloo{$M_{100}$}
\newcommand\Mslll{$M^{sat}_{H \parallel [111]}$}
\newcommand\Msllo{$M^{sat}_{H \parallel [110]}$}
\newcommand\Msloo{$M^{sat}_{H \parallel [100]}$}
\newcommand\DTO{Dy$_2$Ti$_2$O$_7$}
\newcommand\HTO{Ho$_2$Ti$_2$O$_7$}
\newcommand\PRB[3]{Phys. Rev. B {\bf {#1}}, {#2} ({#3})}
\newcommand\PRL[3]{Phys. Rev. Lett. {\bf {#1}}, {#2} ({#3})}
\newcommand\JPCM[3]{J. Phys.: Condens. Matter {\bf {#1}}, {#2} ({#3})}

\begin{document}
\title{Comment on ``Short-range magnetic interactions in the spin-ice compound \HTO"}
\author{O.~A.~Petrenko, M.~R.~Lees and G.~Balakrishnan}
\affiliation{University of Warwick, Department of Physics, Coventry, CV4~7AL, UK}
\date{\today}
\begin{abstract}
In their recent communication (\PRB{64}{060406(R)}{2001}) Cornelius and Gardner have reported the results of magnetization and susceptibility studies on single crystals of the \pyr\ \HTO. The observed field dependence of magnetization is totally unexpected, as it seems to suggest that the magnetic moments in this compound do not obey the 'ice-rule'. We have re-measured the magnetization curves for \HTO\ single crystal for the three principal directions of an applied magnetic field and found nearly perfect agreement with the predictions for a nearest-neighbor spin-ice model.
\end{abstract}

\maketitle
The \pyr\ compound \HTO\ is considered to be a good realization of the spin-ice model, where the magnetic \Ho\ ions possess a strong Ising anisotropy and are constrained to point along the local $\langle 111 \rangle$ direction~\cite{Bramwell_Science01}. This assertion is most vividly corroborated by the elastic neutron scattering patterns obtained in \HTO\ at sufficiently low temperature~\cite{Bramwell}. In the spin-ice model, the following values of saturation magnetization are predicted~\cite{Harris,Siddharthan,Fukazawa}: \Msloo=$\mu/\sqrt{3}$, \Mslll=$\mu/2$ and \Msllo=$\mu/\sqrt{6}$, where $\mu$ is a full magnetic moment of a \Ho\ ion (the corresponding spin configurations are shown in Fig. 1 of Ref.~\onlinecite{Fukazawa}).

Cornelius and Gardner~\cite{Cornelius} have measured the magnetization and susceptibility of single crystals of \HTO. 
Although the authors claimed that the observed magnetization isotherms are qualitatively similar to that predicted by the spin-ice model, two major contradictions with the theoretical predictions are obvious. The first discrepancy is that the saturation magnetic moment~\cite{saturation} of 5.9 $\mu_B$ per Ho atom is reported to be nearly independent of the orientation of magnetic field, while in the spin-ice model \Mlll, \Mllo\ and \Mloo\  must differ. The second discrepancy is that for $H\parallel [110]$ the magnetization is claimed to be almost temperature independent below $T=4$~K.

These are rather surprising results, taking into account that the magnetization curves of a similar spin-ice compound, \DTO, do follow the theoretical predictions quite precisely \cite{Fukazawa,Matsuhira}. In an attempt to clarify the situation we have re-measured the magnetization curves of \HTO\ for the three principal directions of an applied magnetic field, $[100]$, $[110]$ and $[111]$.

Single crystals of \HTO\ have been grown by the floating zone technique using an infrared image furnace \cite{Balakrishnan}. By measuring the X-ray diffraction patterns on powder samples prepared from single crystals, we confirmed the high purity of the crystals, as no impurity peaks were found. The magnetisation versus field data were collected using an Oxford  Instruments vibrating sample magnetometer between 1.6 and 18~K in applied fields of up to 12~T. We have used small (5 to 25 mg) samples of various shapes~\cite{Nfactor}. The absolute accuracy of the magnetization measurements was of the order of 3\% \cite{accuracy}. The principal axes were determined using X-ray diffraction Laue photographs; the crystals were aligned to within an accuracy of 1-2$^{\circ}$.

\begin{figure}
\includegraphics[width=0.93\columnwidth]{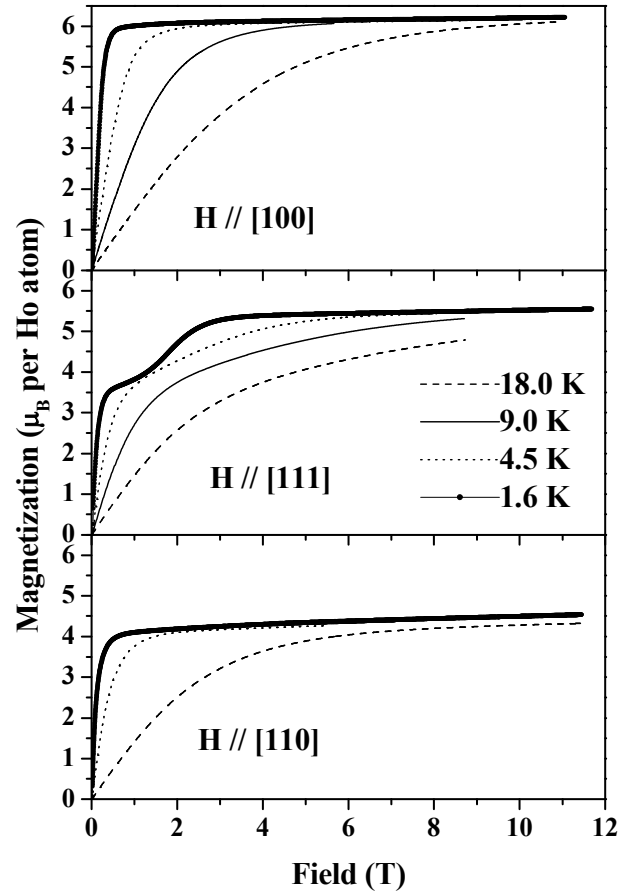}
\caption{\label{Fig1} Field dependence of the magnetization of \HTO\ single crystals at different temperatures.}
\end{figure}

The results of the measurements are summarized in Fig.~1, where the observed magnetization is plotted against the effective magnetic field, $H=H_{ext}-nM$, $n$ being the demagnetization factor \cite{Nfactor}. For the magnetic field applied along the easy-axis direction, $H\parallel [100]$, the magnetization curves look rather similar to those reported by Cornelius and Gardner \cite{Cornelius}. The only small difference is in the value of saturation magnetization: $(6.22 \pm 0.19)\mu_B$/Ho compared to $5.9 \mu_B$/Ho in Ref. \onlinecite{Cornelius}, which could be attributed to the limited experimental accuracy. For $T>2$~K Cornelius and Gardner \cite{Cornelius} obtained identical results for \Mlll\ and \Mloo, while our measurements revealed significantly different values for different orientations at all temperatures. At sufficiently low temperature, $T<2$~K, we observed a plateau in \Mlll\ at $3.6 \mu_B$/Ho in agreement with ref.~\cite{Cornelius}. The obtained value of saturation magnetization, \Mslll, is however rather different: $(5.54 \pm 0.17)\mu_B$ per Ho atom. The saturation is reached in a field of about 3 to 4 T compared to 7 to 8 T reported previously \cite{Cornelius}. Despite the metastable character of the plateau, no significant hysteresis in the magnetisation was detected at any temperature. Note, that the observed ratio \Mslll /\Msloo =$0.89\pm 0.05$ agrees well with the theoretical ratio $\frac{\sqrt{3}}{2} \approx 0.866$. 

Out results for a hard-axis magnetization, \Mllo, are in the sharpest contrast with Cornelius and Gardner report \cite{Cornelius}.
The low-field magnetization ($H<0.3$~T) is found to be temperature dependent and similar to that for the other two orientations of magnetic field. With increasing field, the magnetization grows rapidly and reaches a saturation in a field of about 1.5~T. The value of the saturation magnetization for $H\parallel [110]$ is $(4.43 \pm 0.13)\mu_B$ per Ho atom. Again, the observed ration \Msllo /\Msloo =$0.73\pm 0.04$ complies with the spin-ice theory predicting $\frac{1}{\sqrt{2}} \approx 0.707$. 

From the values of the saturation magnetization for all the three orientations and from the value of \Mlll\ on it plateau, a full magnetic moment on the Ho site is $\mu = (10.9\pm 0.3)\mu_B$. This value is consistent with the free ion moment for the $^5$I$_8$ state of \Ho,  $\mu = 10.6\mu_B$, but is also close to the value expected for an $m_J = \pm 8$ doublet ground state, $\mu = 10.0 \mu_B$ (Ref.~\onlinecite{Bramwell_JPCM}).

The lowest temperature available for our measurements, $T=1.6$~K, is still relatively high on the scale of the exchange interactions in \HTO, therefore we were unable to see the multiple plateaux in magnetization predicted by the Monte Carlo calculations \cite{Siddharthan,Shastry} and observed in the spin-ice compound, \DTO\ (Ref.~\onlinecite{Fennell}).

One can only speculate why the results of Cornelius and Gardner~\cite{Cornelius} are different from our results. Their data cannot be explained solely by a small crystal misalignment, as pointed out in Ref.~\onlinecite{Fukazawa}. An important note to make is that due to a significant anisotropy and the large magnetic moments involved in a system, it is extremely difficult to keep the $H \parallel [110]$ orientation for a sample in a high applied field~\cite{rotation}. One possible explanation is that both for $H \parallel [111]$ and $H \parallel [110]$ orientations the crystal was allowed to rotate in field and eventually aligned itself with the $[100]$ easy-axis parallel to the field. This effect, however, does not explain why the low-field susceptibility for $H\parallel [110]$ is very different from the other two directions~\cite{cube} and why it is almost temperature independent below $T=4$~K. Another possibility is that the actual sample temperature was much higher than stated. Judging by the low-field value of \Mllo, the real temperature of the sample could have been above 20~K.

To conclude, the observed field dependence of magnetization for a single crystal of \HTO\ matches well the predictions for the spin-ice nearest-neighbor model contrary to the report by Cornelius and Gardner~\cite{Cornelius}.


\begin{thebibliography}{99}

\bibitem{Bramwell_Science01} S.~T.~Bramwell and M.~J.~P.~Gingras, Science {\bf 294}, 1495 (2001). 

\bibitem{Bramwell} S.~T.~Bramwell, M.~J.~Harris, B.~C. den Hertog, M.~J.~P.~Gingras, J.~S.~Gardner, D.~F.~McMorrow, A.~R.~Wildes, A.~L.~Cornelius, J.~D.~M.~Champion, R.~G.~Melko and T.~Fennell, \PRL{87}{047205}{2001}.

\bibitem{Harris} M.~J.~Harris, S.~T.~Bramwell, P.~C.~W.~Holdsworth and J.~D.~M.~Champion, \PRL{81}{4496}{1998}.

\bibitem{Siddharthan} R.~Siddharthan, B.~S.~Shastry and A.~P.~Ramirez, \PRB{63}{184412}{2001}.

\bibitem{Fukazawa}  H.~Fukazawa, R.~G.~Melko, R.~Higashinaka, Y.~Maeno and M.~J.~P.~Gingras,
             \PRB{65}{054410}{2002}.

\bibitem{Cornelius} A.~L.~Cornelius and J.~S.~Gardner, \PRB{64}{060406(R)}{2001}.

\bibitem{saturation} The field of 8-9~T where the magnetization curves in Ref.~\onlinecite{Cornelius} appear to reach their maximum value cannot be labelled as a saturation field, because the true ferromagnetic alignment of all the spins is expected to be established only in much higher fields.

\bibitem{Matsuhira} K.~Matsuhira, Z.~Hiroi, T.~Tayama, S.~Takagi and T.~Sakakibara, \JPCM{14}{L559}{2002}.

\bibitem{Balakrishnan} G.~Balakrishnan, O.~A.~Petrenko, M.~R.~Lees and D.~$\rm M^cK$~Paul,
              \JPCM{10}{L723}{1998}.

\bibitem{Nfactor} The estimated demagnetization factor, $n$, varied from 0.3 to 0.65 for different crystals.

\bibitem{accuracy} Although the VSM used could potentially be calibrated to better than 1\% accuracy for a particular size and shape sample, the actual reproducibility of the measurements for different samples is only about 3\%. The background signal from the sample holder was negligible (less than 0.5\%).

\bibitem{Bramwell_JPCM} S.~T.~Bramwell, M.~N.~Field, M.~J.~Harris and I.~P.~Parkin, \JPCM{12}{483}{2000}.

\bibitem{Shastry} B.~S.~Shastry, cond-mat/0210230 (unpublished).

\bibitem{Fennell} T.~Fennell, O.~A.~Petrenko, S.~T.~Bramwell, J.~D.~M.~Champion, B. F{\aa}k, M.~J.~Harris and
        D.~$\rm M^cK$~Paul, to appear in Appl. Phys. A. Material Science \& Processing (2002).

\bibitem{rotation} For a crystal in an unfavorable orientation, even in a moderate field of 3 - 4 T the magnetic forces involved are strong enough to break the crystal itself, not to mention the adhesive materials. Special precautions need to be taken to ensure that the crystal does not move in an applied field.

\bibitem{cube} This observation is in contradiction with the cubic symmetry of the crystal.

\end{thebibliography}
\end{document}